\documentclass[12pt,a4paper]{iopart} 

\usepackage{iopams}  
\usepackage{bm}
\usepackage{enumerate} 
\usepackage{graphicx}
\usepackage{rotating}
\usepackage[english,activeacute]{babel}
\usepackage{amssymb, latexsym}
\usepackage{flafter}
\usepackage{upgreek}
\usepackage{color}

\usepackage{cite}
\begin{document} 
\title{RF dressed atoms beyond the linear Zeeman effect}
\author{G. Sinuco-Le\'on and B.M. Garraway} 
\address{Department of Physics and Astronomy, University of Sussex, Falmer, Brighton, BN1 9QH, United Kingdom}
\date{\today.}
\begin{abstract}
  We evaluate the impact that non-linear Zeeman shifts have on resonant
  radio-frequency dressed traps in an atom chip configuration. The
  degeneracy of the resonance between Zeeman levels is lifted at large
  intensities of a static field, modifying the spatial dependence of the
  atomic adiabatic potential. In this context we find effects which are
  important for the next generation of atom-chips with tight trapping: in
  particular that the vibrational frequency of the atom trap is sensitive to
  the RF frequency and, depending on the sign of the Land\'e factor, can
  produce significantly weaker, or tighter trapping when compared to the
  linear regime of the Zeeman effect. We take  $^{87}$Rb as an example and
  find that it is possible for the trapping frequency on $F=1$ to exceed
  that of the $F=2$ hyperfine manifold.
 \end{abstract}

\submitto{\NJP}
\maketitle

\section{Introduction}

The use of radio-frequency fields (RF) for the manipulation of ultra-cold
atomic samples
\cite{PhysRevLett.86.1195,PhysRevA.69.023605,0295-5075-67-4-593,NP1_57,PhysRevA.73.033619,PhysRevA.74.053413}
is underpinning important developments in areas such as matter-wave
interferometry
\cite{NP1_57,PhysRevLett.98.030407,0953-4075-43-5-051003,PhysRevLett.105.243003,PhysRevLett.105.265302,RevModPhys.81.1051}
which are beyond the original use of RF fields for evaporative cooling
\cite{LaserCoolingBook}.  Nowadays, in combination with atom-chip technology
\cite{AtomChipsBook} or optical lattices
\cite{PhysRevLett.100.150401,PhysRevA.78.051602} RF dressing is a well
established technique that allows us routinely to control and engineer
atomic quantum states and potential landscapes on micron scales
\cite{RMP79_235,AtomChipsBook}.  Furthermore, RF dressing plays a central
role in several proposals for extending the scope of functions and
applications of ultra-cold atomic gases, including reduced dimensionality
and connected geometries (ring and toroidal traps)
\cite{PhysRevA.73.033619,PhysRevLett.99.083001,PhysRevA.75.063406,0295-5075-67-4-593,1367-2630-10-4-043012,
  PhysRevA.85.053401,PhysRevA.81.031402,PhysRevA.83.043408}, cooling and
probing of RF dressed atom traps
\cite{morizot_trapping_2007,PhysRevA.74.053413,0953-4075-43-6-065302},
sub-wavelength tailoring of potentials \cite{PhysRevLett.100.150401} and
transporting atoms in dressed atom traps \cite{morgan_coherent_2011}.

Experimental realizations of RF dressed magnetic traps have worked in a
range of static field intensities that produce linear Zeeman energy shifts
\cite{0295-5075-67-4-593,PhysRevA.74.023616,morizot_trapping_2007,epjde2008_00050-2,AtomChipsBook,PhysRevA.74.023617}.
Nevertheless, developments in near surface trapping/control and
micro-fabrication \cite{Folman_MatSciRev} indicate that production of strong
trapping configurations and sub-micron control will be soon on the agenda
\cite{nanomagnetictrap,PhysRevA.83.021401} and a full description of the
atomic Zeeman shifts becomes relevant.

One impressive application of RF dressing of magnetic traps is the
miniaturized matter-wave interferometer for coherent spatial splitting
and subsequent stable interference of matter waves on an atom chip
\cite{NP1_57,RevModPhys.81.1051,PhysRevLett.106.025302}. In the
interferometer, the potential landscape typically comprises a double
well in a transverse plane, accompanied by longitudinal weak trapping
\cite{NP1_57,PhysRevLett.106.025302, PhysRevA.77.063623,
  PhysRevA.76.013401}. It has been proposed that the double well
potential can be helpful to study entanglement and squeezing
phenomena, and to study phase coherence dynamics and many-body
quantum physics \cite{PhysRevLett.106.025302,PhysRevA.81.043621}. Here
we establish the relevance of non-linear Zeeman shifts on the
production of a double well potential in a typical atom chip
configuration.

\begin{figure}[!htb]
\centering
\includegraphics[width=4.2cm]{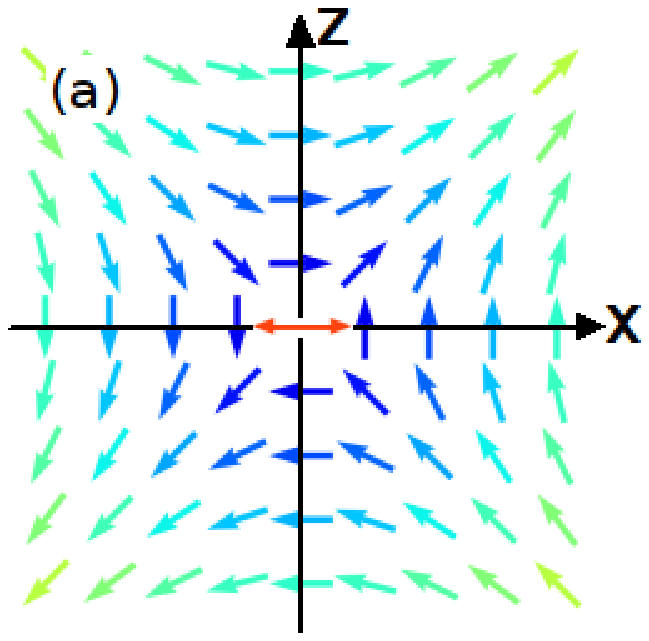}
\includegraphics[width=5.0cm]{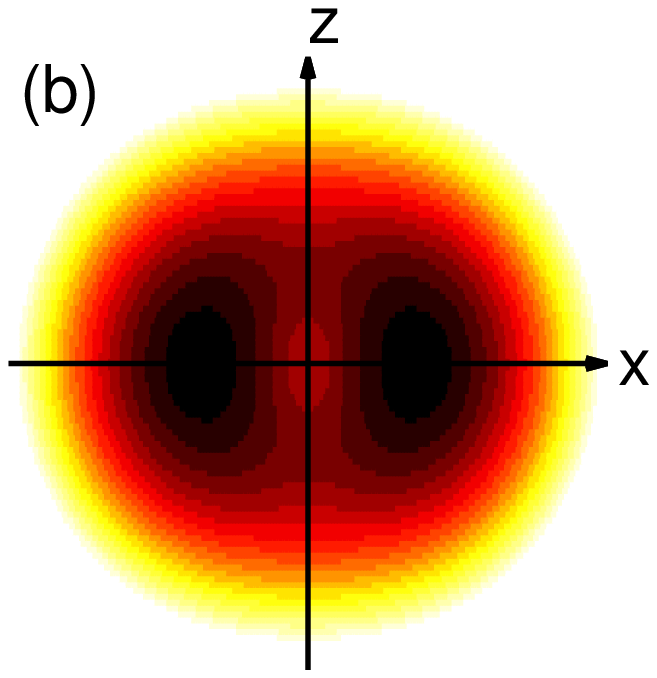}
\includegraphics[width=6.2cm]{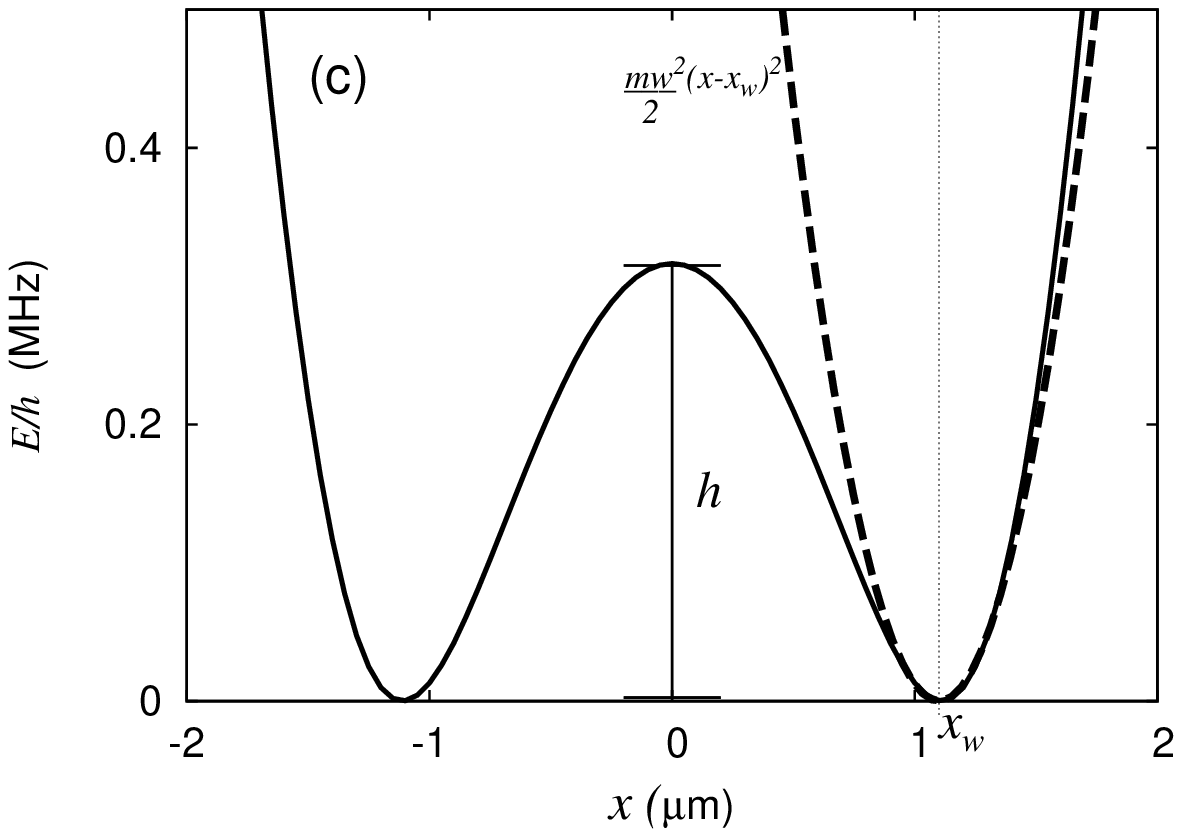}
\caption{\label{quadrupoleField} (a) Schematic of a magnetic field
  quadrupole distribution [equation (\ref{eq:StaticField})] and the RF coupling field.
  The horizontal double-headed arrow (red) indicates the polarization of the
  coupling RF field. The offset field $B_0$ points out of the page.  (b)
  Contours of a typical RF dressed double well potential. (c) Potential
  energy along $y=0$ (the $x$-axis) in panel (b).  Parameters characterizing
  the double well are defined as: well position $x_w$, inter-well barrier
  height $h$ and well frequency $\omega$.}
\end{figure}

For the purposes of this work, the potential landscape is produced by
resonantly dressing a static magnetic field comprised of a quadrupole field with an
offset field. Then the overall field components are
\begin{equation}
  \boldsymbol{B}_{DC} = (G\,y,G\,x,B_{0})   \,.
\label{eq:StaticField}
\end{equation} 
Here $G$ is the magnetic field gradient near the quadrupole centre and
$B_0$ is the uniform offset field (see figure \ref{quadrupoleField}).
In order to give quantitative results we focus on the ground state
manifold of $^{87}$Rb as an example, and evaluate the dressed
potentials for the $|F=2,m_F=2\rangle$ state and also give some
results for the $|F=1,m_F=-1\rangle$ state.  For simplicity, our
analysis is restricted to small amplitudes of the RF field ($B_{RF}$),
such that beyond rotating wave approximation (RWA), effects can be
ignored \cite{PhysRevA.76.013401}. This is also helpful to identify
clearly the effects due to non-linearity of the energy shifts. To
investigate the relevance in current experimental situations, we took
parameters from recent experiments, i.e.\ we took $G=22.67$ T/m,
$B_0=1.0$ or $ 3.0$ G and $B_{RF} = 357$ mG   
\cite{PhysRevA.76.013401}: this enables us to quantify the effect that
the non-linear Zeeman shifts have on the shape of dressed potential.

In the following we first review the theory of the Zeeman effect in
non-linear situations (section \ref{sec:zeeman}) and then apply the results
to RF dressing in the strong field regime (section \ref{sec:strongfield}).
We report on the effects on both the vibrational frequency of RF dressed
atom traps and their locations, and then the paper concludes in section
\ref{sec:conc}.

\section{Zeeman shift theory}
\label{sec:zeeman}

In the limit of slow atomic motion, the magnetic moment of the
atoms keeps its orientation relative to a spatially varying magnetic
field, $\boldsymbol{B}_{DC}$. The dynamics can be described by the
Hamiltonian \cite{AlanCorney,FootBook}
\begin{equation}
  H = A \boldsymbol{I}\cdot \boldsymbol{J} + \mu_B|\boldsymbol{B}_{DC}|(g_J J_z - g_I' I_z) ,
\label{ZeemanHamiltonian}
\end{equation}
where $A$ is a measure of the hyperfine splitting, $\boldsymbol{J}$ and
$\boldsymbol{I}$ are the spin and nuclear angular momentum operators, and
the electronic and nuclear $g$-factors are $g_J$ and $g_I'$, respectively.
The Bohr magneton is denoted by $\mu_B$, as usual.

As is well known, the operator $F_z=J_z+I_z$ commutes with the Hamiltonian
(\ref{ZeemanHamiltonian}) and thus $m_F$ is a good quantum number (along
with $I$ and $J$).  Since we consider $^{87}$Rb
with $I=3/2$ and $J=1/2$, the interaction mixes pairs of
states $|m_F-1/2,1/2\rangle$ and $|m_F+1/2,-1/2\rangle$ in a $m_I$, $m_J$ basis:
$|m_I,m_J\rangle$.
These two states have the same value of $m_F$ by construction.
In this case, because $J=1/2$, the diagonalization of 
Hamiltonian (\ref{ZeemanHamiltonian}) reduces to $2 \times 2$ matrix blocks
and some uncoupled terms. This 
leads to a Breit-Rabi formula for the hyperfine energy spectrum of an
alkali atom in a magnetic field \cite{AlanCorney}: 
\begin{eqnarray}
E^{Z,\pm}_{m_F}&=& -\frac{A(1+4\alpha g_I' m_F)}{4} 
\nonumber\\ &&
\pm \frac{A}{2}\sqrt{ I(I+1) + \frac{1}{4} + 2 m_F \alpha (g_J+g_I') + \alpha^2(g_J+g_I)^2}
\label{Zeeman_energy}
\end{eqnarray}
where $\alpha= \mu_B |\boldsymbol{B}_{DC}|/A$.  However, there are two
uncoupled states in the $m_I$, $m_J$ basis: $|I,1/2\rangle$ and
$|-I,-1/2\rangle$. These states have energies
\begin{eqnarray}
E^{Z,\pm}_{m_F=\pm(I+1/2)}&=& \frac{A}{2} I  \pm \frac{\alpha A}{2}\left( g_J -  2g_I'I \right)  \,.
\label{Zeeman_energy_special}
\end{eqnarray}

\begin{figure}[!htb]
\centering
\includegraphics[width=15.0cm]{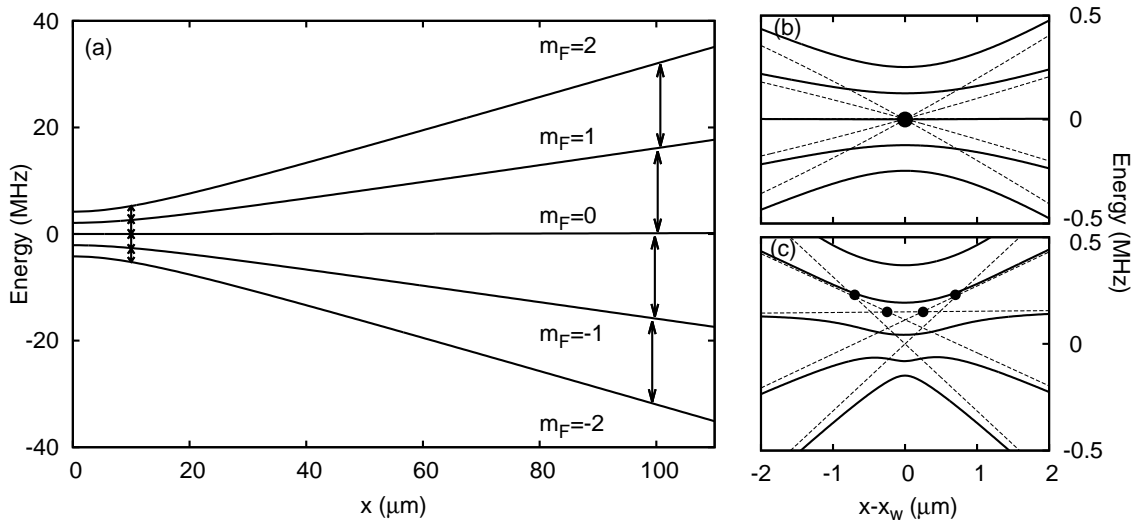}
\caption{\label{fig:ZeemanShifts} (a) Zeeman shifts for the upper manifold ($F=2$)  
of the ground state of $^{87}$Rb, in a field distribution as equation 
  (\ref{eq:StaticField}). Arrows indicate locations of resonance at low
  (left) and high (right) field intensities. Panels (b) and (c) show dressed
  potentials (solid) and bare states (dashed) for resonance located at (b)
  $x_w = 10 \mu$m, with $\omega_{RF} / 2 \pi = 2.63$ MHz and 
  (c) $x_w = 100  \mu$m with $\omega_{RF}/ 2 \pi = 15.99$ MHz. 
  Notice how, at large fields, the 
  crossings of bare states have separated in space and in energy (dashed
  lines). 
  In all cases the field gradient is $G=22.67$ T/m, the RF amplitude is $B_{RF} = 357$ mG,  and the offset field is $B_0 = 3$ G.  }
\end{figure}

In the \emph{weak field regime}, corresponding to $\alpha \ll 1$, the Zeeman
shifts become linear in $|\boldsymbol{B}_{DC}|$ and all these
energy levels [equation (\ref{Zeeman_energy}) and equation
(\ref{Zeeman_energy_special})] are approximated by (see e.g.\ Ref.\ \cite{FootBook})
\begin{eqnarray}
  E^{Z,\pm}_{m_F} &\approx& -\frac{A}{4} + \frac{A}{2} (I + 1/2) + m_F \mu_B g_F|\boldsymbol{B}_{DC}|,
\, .
\label{eq:linealZeeman}
\end{eqnarray}
Here the signs in front of the square root contribution in equation
(\ref{Zeeman_energy}) have been absorbed by the definition of the $g$-factor
associated with the total angular momentum
$\boldsymbol{F}=\boldsymbol{I}+\boldsymbol{J}$. Because $J=1/2$ there are
only two possible values $F=I+J=2$ and $F=|I-J|=1$:
\begin{eqnarray}
g_F&=& g_J \frac{F(F+1) - I(I+1) + J(J+1)}{2F(F+1)} 
\nonumber \\&&
- g_I'\frac{F(F+1) + I(I+1) - J(J+1)}{2F(F+1)}
\,.
\label{eq:gf}
\end{eqnarray}
Figure \ref{fig:ZeemanShifts}(a) shows the atomic energy levels,
corresponding to the upper hyperfine manifold, at different positions in the
field distribution equation (\ref{eq:StaticField}). In the case of RF dressed
atom traps with $^{87}$Rb atoms in the trapping states $m_F=1,2$ follow such
energy curves adiabatically and become trapped at positions of minimum field
amplitude \cite{RMP79_235}.

\section{Double well dressed potential}
\label{sec:strongfield}

We consider a magnetic trap with static field distribution
(\ref{eq:StaticField}), dressed by a uniform and linearly polarized
oscillating magnetic field (RF) \cite{NP1_57,PhysRevLett.86.1195}. The
coupling between atomic states is dominated by the component of the RF field
orthogonal to $\boldsymbol{B}_{DC}$ \cite{PhysRevA.73.033619}, vanishing at
positions where they are parallel (e.g.\ at positions of coordinates
$(x=0,y)$).  A typical resulting double well potential is shown in figure
\ref{quadrupoleField}(b). To characterize the potential energy, we evaluate
the dressed energy along the $x$ axis, where a double well appears as
a consequence of the dressing.

\begin{figure}[!htb]
\centering
\includegraphics[width = 12.0cm]{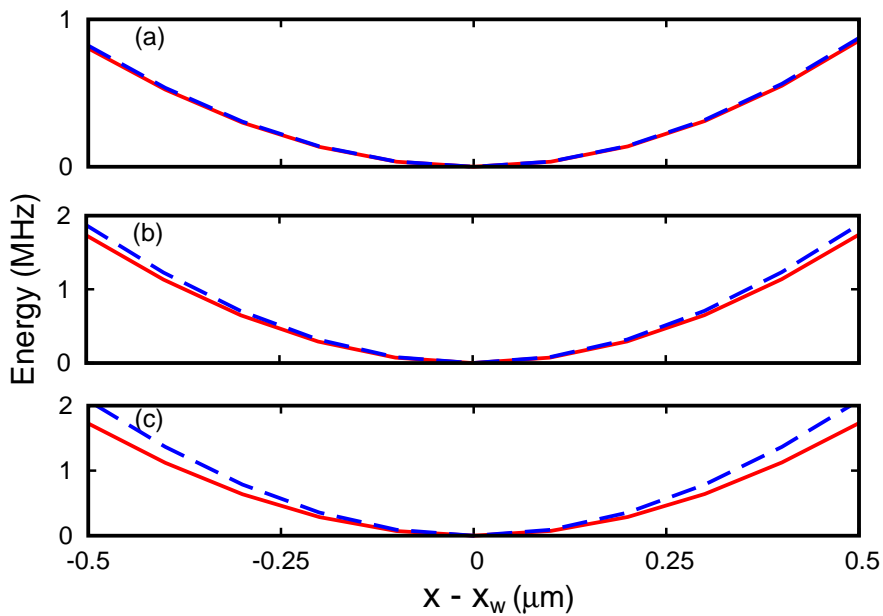}
\caption{\label{doubleWell_}Comparison of non-linear and linear dressed potentials.
  There are two potential wells at $\pm x_w$ and we show here just the
  well located at $x=+x_w$ with $m_F=2$: Red (solid) eigenvalue of
  Hamiltonian 
  (\ref{RWA}).  Blue (dashed) line is the adiabatic potential
  (\ref{V_adb}).  The dressing frequencies and well positions are (a)
  $2.6$ MHz, $x_w = 10 \ \mu$m (b) $5.2$ MHz, $30 \ \mu$m and (c)
  $8.2$ MHz, $50 \ \mu$m.  The quadrupole field gradient is $G=22.67$
  T/m \cite{PhysRevA.76.013401} and the bias field is $B_0 = 3.0$ G as
  in figure \ref{fig:ZeemanShifts}.  
  In each of (a-c), $\omega_{RF}$ has
  been adjusted to produce a minimum at the specified value of $x_w$.}
\end{figure}

We define the $z$-axis parallel to direction of the static field, take
the RF field polarized along the $x$-axis and apply the standard
rotating wave approximation (RWA). Then, the Hamiltonian
(\ref{ZeemanHamiltonian}) in a rotating
frame becomes
\begin{equation}
  H = A \boldsymbol{I}\cdot \boldsymbol{J} + \mu_B|\boldsymbol{B}_{DC}|(g_JJ_z - g_I' I_z) \pm \hbar \omega_{RF} (J_z+I_z) + \frac{\mu_B |\boldsymbol{B}_{RF}|}{2} (g_J J_x - g_I' I_x) .
\label{RWA}
\end{equation}
Because $J=1/2$ the sign $\pm$ is chosen according to which hyperfine
manifold we are interested in: $F=I+J$ or $F=|I-J|$. This is because each
polarization component of the dressing field couples magnetic sublevels
within a subspace with a given total angular momentum (\ref{sec:app-brf}). The
diagonalization of equation (\ref{RWA}) produces the dressed state energies
displayed in figure \ref{doubleWell_} (solid curve) for the magnetic field
configuration of this paper.

In the weak field regime and for RF frequencies much smaller than the
hyperfine splitting ($\omega_{RF} \ll A/\hbar$), the dressed energies for
states of the upper hyperfine manifold are \cite{PhysRevLett.86.1195}
\begin{equation}
E^+_{m_F} \approx -\frac{A}{4} + \frac{A}{2}(I+1/2) + m_F \sqrt{(\mu_B g_F|\boldsymbol{B}_{DC}| - \hbar
  \omega_{RF})^2 + (\mu_B g_F|\boldsymbol{B}_{RF}|/2)^2} ,
\label{V_adb}
\end{equation}
which produces, for the field distribution of equation (\ref{eq:StaticField})
and for weak-field seeker states \cite{RMP79_235}, the dressed potentials shown
in figure \ref{doubleWell_} (dashed line). The double well is conveniently
described by three parameters: the well minimum position ($x_{w}$), the
height of the inter-well barrier ($h$) and the well frequency $\omega$ which is defined
through a harmonic approximation centred around the potential minimum [see
figure \ref{quadrupoleField}(c)].  To distinguish these properties being
evaluated according to equation (\ref{V_adb}) or equation (\ref{RWA}), we denote the
former quantities (i.e.\ from the weak field expression) with a 
superscript $0$. (For example, the well position is $x_w^0$ in the linear
regime, and $x_w$ more genrally.)

The potential well's position and frequency are determined by the location
of the resonant coupling, as suggested in figure \ref{fig:ZeemanShifts}(b)-(c),
and by the intensity of the RF field.  Figure \ref{doubleWell_} shows 
typical dressed potentials in a double-well regime (where we show only one
of the two wells). 
We compare both the dressed eigenvalue of
Hamiltonian (\ref{RWA}), and equation (\ref{V_adb}) for different RF
frequencies $\omega_{RF}$. A rather significant change in the well shape is seen
for quite modest well separations for this field gradient. (The well
separation is $2x_w$ in figure \ref{doubleWell_}.)

\begin{figure}
\centering
\includegraphics[width=15.0cm]{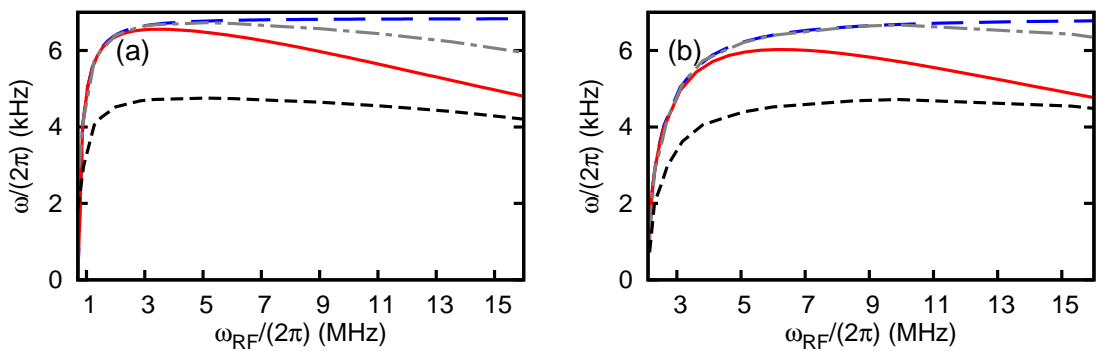}
\caption{\label{w_trap_shiftsF2} Well frequency as a function of RF frequency,
  obtained from the eigenvalue of Hamiltonian (\ref{RWA}) (solid) and
  the linear approximation to the eigenvalues (\ref{V_adb}) (dashed),
  for (a) $B_0=1.0$ G and (b) $B_0=3.0$. Other parameters are as in figure
  \ref{fig:ZeemanShifts} (including $F=2$, $m_F=2$). The well 
  frequency corresponding to a three level system and its value scaled by $\sqrt{2}$ are shown by the short-dash and
  dot-dash lines, respectively.   
}
\end{figure}

With a fixed static field configuration, the distance between the wells is
controlled by the RF frequency $\omega_{RF}$. At large distances, the non-linear
character of Zeeman shifts impacts on the dressed potential, and equation
(\ref{V_adb}) is no longer valid. In particular, the well frequency is
significantly modified, as seen for $m_F=2$ in figure \ref{w_trap_shiftsF2}.  For well
separations of 30 microns, the correction to frequency
is about $5\%$ for offset fields of $1$ G (or $3$ G).
The relative effect on the frequency increases approximately
linearly with the well separation, and can become a significant fraction of
$w$ for separations of tens of microns. The weakening of the well (or
equivalently, the reduction of $w$), can be qualitatively understood as a
consequence of the appearance of multiple resonant positions due to
non-linearity of Zeeman shifts [see figure \ref{fig:ZeemanShifts}(b)]: as
the distance between bare state crossings increases, the dressed potential
softens in comparison with the potential from a single resonant position.
This becomes one of the main effects of the non-linear Zeeman effect
on dressed RF potentials.

\begin{figure}
\centering
\includegraphics[width=15.0cm]{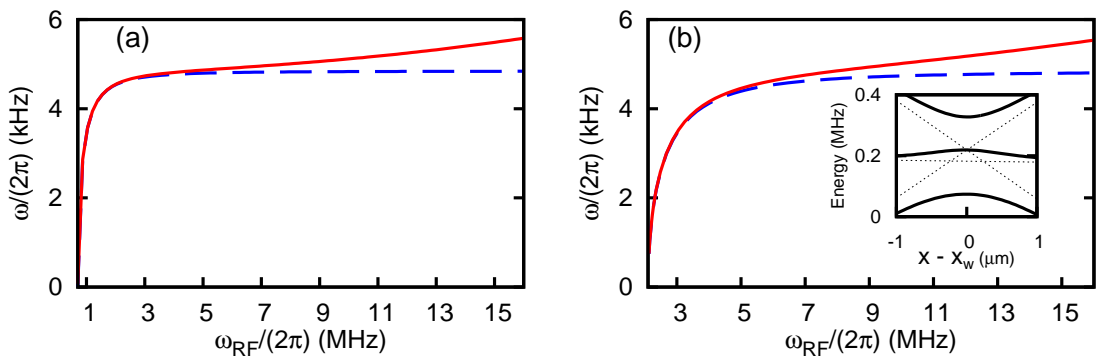}
\caption{\label{w_trap_shiftsF1} Well frequency as a function of RF
  frequency in the case $F=1$, $m_F=-1$. Results are obtained from the
  eigenvalue of Hamiltonian  (\ref{RWA}) (solid) and the linear
  approximation to the eigenvalues  (\ref{V_adb}) (dashed), for (a)
  $B_0=1.0$ G and (b) $B_0=3.0$. Other parameters are as in figure
  \ref{fig:ZeemanShifts}.
   Inset: dressed potentials for the case of $F=1$.
   }
\end{figure}

Figure \ref{w_trap_shiftsF1} shows similar results for what is the $F=1$
manifold in the weak field regime. However, in this case we see that the
trap is \emph{tightened} by the non-linear Zeeman effect. The result is
understood from the fact that $g_F$, equation (\ref{eq:gf}), changes sign for
$F=1$. As a result the sequence of levels seen for $F=2$ in figure
\ref{fig:ZeemanShifts}(c) is reversed, which enables us to have tighter
trapping as seen (inverted) at the bottom of the manifold of figure
\ref{fig:ZeemanShifts}(c)
(and see also figure \ref{w_trap_shiftsF1}(b) inset).
Since the trapping on $^{87}$Rb $F=2$ involves
potentials twice as steep as those on $^{87}$Rb $F=1$ we would not expect the
trap to become tighter in $F=1$ overall. (The factor two in steepness is
because the two manifolds have the same magnitude of $g_F$, which means a
factor of two in the potentials because for $F=2$ we can use $m_F=2$, while
for $F=1$ we can only use $m_F=-1$.) However, figures \ref{w_trap_shiftsF2}
and \ref{w_trap_shiftsF1} show that, in absolute terms, that $F=1$ trap does
become tighter than the $F=2$ trap for quite modest values of the RF
frequency. In essence, the weakening of the $F=2$ trap by the non-linear
Zeeman effect, and corresponding tightening of $F=1$, becomes sufficient for
the trap frequency in $F=1$ to become higher.

The potential well's frequency for the dressed state $|F=1,m_F=-1\rangle$ can be evaluated analytically using the solution of a three level system (see \ref{sec:app-position}), and approximating the well's position by taking the positions of the resonant couplings (large dots in figure \ref{fig:ZeemanShifts}(c)) and averaging them. The frequency obtained following this procedure coincides very closely with the numerical results shown in figure 5.

In the case of $F=2$, we find that, similarly, the well frequency corresponding to the dressed state $|F=2,m_F=2\rangle$ can be estimated by considering couplings between the bare levels $|F=2,m_F=-1\rangle,|F=2,m_F=0\rangle$ and $|F=2,m_F=1\rangle$. This is shown in figure \ref{w_trap_shiftsF2} (short-dash) and indicates that at high RF frequencies the curvature of the dressed energy is dominated by the crossings of just these three levels  (see figure \ref{fig:ZeemanShifts}(c)). In the intermediate regime of dressing frequency, the coupling of all levels contribute significantly to the dressed state and the three level approximation is not valid. However, at low RF dressing frequencies,  we observe that the linear regime result coincides with the three level estimate of the well frequency scaled by $\sqrt{2}$ (dash and dash-dot line in figure \ref{w_trap_shiftsF2}, respectively).  

\begin{figure}
\centering
\includegraphics[width=15.0cm]{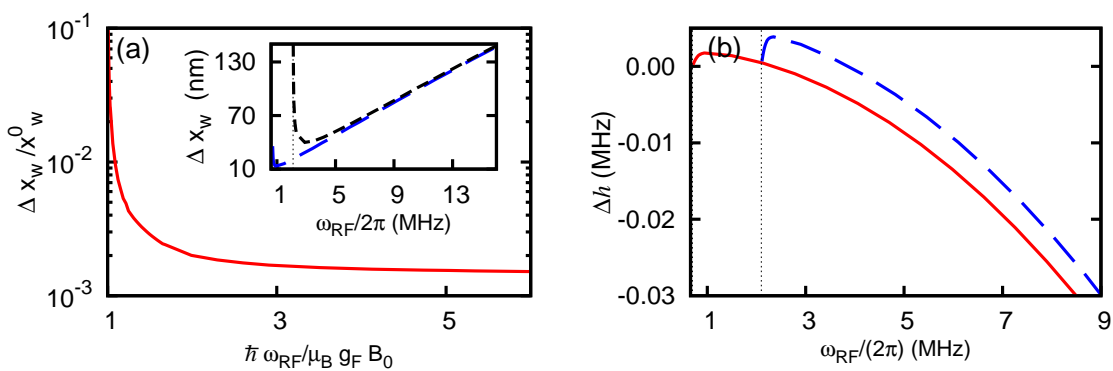}
\caption{\label{DHandDs} (a) Main panel: Relative displacement of well
  positions ($\Delta x_w/x^0_w = (x_w-x^0_w)/x^0_w$) versus $\hbar
  \omega_{RF}/\mu_BB_0$, evaluated via equation (\ref{RWA}). On
  this scale, $\Delta x_w/x^0_w$ are approximately independent of $B_0$. Inset: Well
  position displacement for offset field $B_0=1$ G (dashed) and $B_0=3$ G
  (short-dashed) (b) Modification of the double well barrier height ($\Delta
  h = h-h^0$) for $B_{0}=1$ G (solid) and $B_0=3$ G (dashed) respectively,
  as function of the RF frequency. In all cases $B_{RF} = 357$ mG and
  $G=22.67$ T/m \cite{PhysRevA.76.013401}. Dashed vertical lines indicate
  the limiting frequencies at $\mu_B g_F B_0/h$.}
\end{figure}

Figure \ref{DHandDs} presents a comparison of the non-linear effects
on the well position ($\Delta x_w = x_w-x_w^0$) and inter-well barrier
height ($\Delta h = h-h^0$). Again, these are calculated from the
diagonalization of the Hamiltonian (\ref{RWA}) and by comparing
results to the weak field approximation, equation (\ref{V_adb}). We see
that the shift in the position of a potential well can be large in
both the low RF frequency limit and the high RF frequency limit (see
figure \ref{DHandDs}(a), inset). The shift at high RF frequency seems
reasonable as the higher frequencies usually result in RF resonance in
regions of stronger magnetic fields where the non-linear Zeeman effect
is expected to play a role. Note here that the vertical asymptotes in
figure \ref{DHandDs} correspond to the threshold were the RF frequency is
just sufficient to excite transitions in the static field configuration,
i.e.\  $\hbar \omega_{RF} \rightarrow \mu_B g_F |B_0|$.

To understand the shifts in position of the potential wells 
we look carefully at the issue of how we define the resonance
location and the potential well location.
Shifts of the well positions, $\Delta x_w = (x_w-x^0_w)$, can be understood
by investigating where resonant RF coupling occurs, i.e, positions where
$\hbar \omega_{RF}$ matches the energy separation between levels with $|\Delta
m_F| = 1$. That is, 
\begin{equation}
 E^{Z,+}_{m_F}-E^{Z,+}_{m_F \pm 1}  = \hbar \omega_{RF}
\label{Resonance_condition}
\end{equation}
which, in the weak field regime, reduces to a unique value of $\omega_{RF}$ given
by
\begin{equation}
 \mu_B g_F|\boldsymbol{B}_{DC}| - \hbar \omega_{RF} = 0 .
\label{eq:ResonanceLineal}
\end{equation}

In the regime of weak magnetic field, multiple resonances occur at the same
position [see figure \ref{fig:ZeemanShifts}(b)], since the spacing between
adjacent sub-levels is degenerate. In contrast, for large fields the
degeneracy of transition frequencies is destroyed and the resonance
locations become separated in space, as shown in figure
\ref{fig:ZeemanShifts}(c). 
As in the treatment of the well frequencies above,
a simple and good estimate for the location of the potential well
minimum $x_{w}$ can be found by taking an average of the positions where equation
(\ref{Resonance_condition}) is satisfied for different $(m_F, m_F \pm 1)$
pairs [dots in figure \ref{fig:ZeemanShifts}(c)]. This quantity,
$\bar{x}_w$, has been evaluated using both the procedure of
Appendix B (equation \ref{eq:Br_approx}), and by using the numerical solution of equation
(\ref{Resonance_condition}) itself. These results agree with the full
solution of equation (\ref{RWA}) and are presented as one
curve in figure \ref{DHandDs}.

\section{Conclusions} 
\label{sec:conc}

Our calculations show that for a resonant RF dressed double well potential,
the well frequency is a sensitive parameter to the non-linearity of Zeeman
shifts. This is relevant for investigations of tunnelling processes in
double well potentials, BEC interferometry, and phase coherence dynamics,
since these phenomena are sensitive to the potential shape
\cite{PhysRevLett.106.025302,PhysRevA.81.043621}.  These modifications of
the well frequency can be large enough to be tested experimentally via
interferometry or atom cloud oscillations \cite{0295-5075-67-4-593,NP1_57}.

We have also seen that in taking account of the non-linear Zeeman effect the
sign of the $g$-factor is an important consideration in the tightness of the
resonantly RF dressed atom trap. In the case of $^{87}$Rb we found that for
modest RF frequencies the $F=1$ dressed trap could be tighter than the $F=2$
dressed trap, which is quite against expectation in the linear regime. By using 
a model 3-level system, we calculate the well's frequency for $F=1$. 
In 
the case of the $F=2$ manifold,  this model works well in a regime of non-linear Zeeman shift. 
However, the presence of tighter trapping in $F=1$ seems counter-intuitive,
not least because the key states of $m_F=\pm 1$ do not directly
couple. However, the cases we have examined have sufficiently strong
coupling that all the levels are mixed by the interaction even though the
crossings are seen to be separated in figure \ref{fig:ZeemanShifts}c.

Finally, we note that near the limiting frequency set by the static offset
field (see figure \ref{DHandDs}), the double well separation and frequency
of the dressed potential are quite sensitive to the dressing frequency
$\omega_{RF}$ (see also figures \ref{w_trap_shiftsF2} and \ref{w_trap_shiftsF1}).
This regime is relevant for double wells with sub-micron separations, a
situation likely to occur in near surface trapping configurations 
\cite{nanomagnetictrap}. Our work shows that to achieve stable configurations 
with small well separations, good control of the RF frequency is needed. 
However, the main result of this work is that the vibrational frequency can be 
sensitive to the chosen RF frequency because of the non-linear nature of  Zeeman shifts.

\ack
We thank H{\'e}l{\`e}ne Perrin for useful discussions and 
gratefully acknowledge funding from the EPSRC (grant EP/I010394/1).

\appendix
\section{RF dressing in the strong field regime}
\label{sec:app-brf}
\setcounter{section}{1}

An alkali atom interacting with a static magnetic field plus an orthogonal RF dressing field is described by the Hamiltonian:
\begin{equation}
  H = A \boldsymbol{I}\cdot \boldsymbol{J} + \mu_B|\boldsymbol{B}_{DC}|(g_J J_z - g_I' I_z) + \mu_B|\boldsymbol{B}_{RF}|(g_J J_x - g_I' I_x) \cos(\omega_{RF}t)
\label{ZeemanHamiltonian_appendix}
\end{equation}
where the first term arises from the hyperfine interaction.
Assuming that the hyperfine coupling is stronger than the interaction with
the static field, the state space can be split into a direct sum of spaces
with two angular momentum operators $F^{\downarrow}$ (which has spin $I-J$)
and $F^{\uparrow}$ (which has spin $I+J$). Left and right circular
polarization components of a linearly polarized dressing field couple
magnetic levels within only one of the hyperfine subspaces. Thus, the
relevant component in each case is selected by an observer rotating in an
appropriate sense. With this in mind, the unitary transformation between lab
and rotating frames can be defined by \cite{PhysRevA.85.022302}:
\begin{equation}
U =  \exp(-i\omega_{RF}t(F_z^{\uparrow}-F_z^{\downarrow}))
\,,
\label{eq:rotation_appendix}
\end{equation}
where $F_z^{\uparrow}$ and $F_z^{\downarrow}$ work in subspaces of
$\mathbf{F}$ as described above.
Then, in the rotating frame, the time evolution follows the Schr\"odinger equation $i\hbar \partial_t |\psi\rangle =  H'|\psi\rangle$ with
\begin{equation}
H' = U^\dagger H U - i \hbar U^\dagger \dot{U}.
\end{equation} 
The hyperfine term and the interaction with the static field are invariant under the unitary transformation equation (\ref{eq:rotation_appendix}). The interaction with the dressing field transforms into:
\begin{eqnarray}
H'_{RF} &=& \frac{\mu_B B_{RF,x}}{2}(g_JJ_x-g_I'I_x) \nonumber \\
& & + \frac{\mu_B B_{RF,x}}{2}(\cos 2\omega_{RF}t \ \hat{x} \mp \sin 2\omega_{RF}t \ \hat{y})\cdot(g_J\boldsymbol{J}-g_I'\boldsymbol{I}) .
\end{eqnarray}
Neglecting the rapidly rotating field (rotating with angular velocity of $2
\omega_{RF}$), the total Hamiltonian becomes the expression given in equation (\ref{RWA}).

\section{ Frequency and position of strong-field potential wells}
\label{sec:app-position}

Consider the dressed energy as function of the static magnetic field, as given, e.g.\ in equation (\ref{V_adb}), and produce a Taylor expansion around the minimum occurring at the magnetic field $|\boldsymbol{B}_{DC,min}|$,
\begin{equation}
E(|\boldsymbol{B}_{DC}|) = E_0 + \frac{\beta}{2} (|\boldsymbol{B}_{DC}|-|\boldsymbol{B}_{DC,min}|)^2 + \ldots
\label{eq:Taylor_B}
\end{equation}
with $\beta$,
\begin{equation}
\beta = \left. \frac{d^2E(|\boldsymbol{B}_{DC}|)}{d|\boldsymbol{B}_{DC}|^2} \right|_{|\boldsymbol{B}_{DC,min}|}.
\label{eq:beta}
\end{equation}
The well's frequency is parametrized by a quadratic dependence of the dressed energy with the distance to the position of minimum
\begin{equation}
E(x) = E_0 + \frac{m w^2}{2} (x-x_{w})^2
\label{ea:Taylor_x}
\end{equation}
where
\begin{equation}
|\boldsymbol{B}_{DC,min}| = |\boldsymbol{B}_{DC}(x_{w})|. 
\end{equation}

In the case of a quadrupole  magnetic field of gradient $G$ plus offset field, where the field is $|\boldsymbol{B}_{DC}(x)| =  \sqrt{(Gx)^2 + B_{0}^2}$, the well's frequency in terms of expansion equation (\ref{eq:Taylor_B}) is given by: 
\begin{equation}
w = \sqrt{\frac{\mu_B^2 \beta}{mA}} G \sqrt{1-\frac{B_{0}}{|\boldsymbol{B}_{DC,min}|}}.
\label{eq:w_beta_terms}
\end{equation}
Then in the regime of linear Zeeman shifts, and after a Taylor expansion to second
order of equation (\ref{V_adb}) we obtain
\begin{eqnarray}
|\boldsymbol{B}_{DC,min}| &=& \frac{\hbar \omega_{RF}}{g_F\mu_B} \nonumber \\
\beta &=&  \frac{8 A g_F}{\mu_B |\boldsymbol{B}_{RF}|}  .
\end{eqnarray}
For this to be valid, the minimum should occur at a magnetic field such that $\mu_B |\boldsymbol{B}_{DC,min}| \ll 2A$.

For more intense fields, where the multi-level crossing degeneracy is lifted, but levels can still grouped in manifolds $F=I+J$ and $F=|I-J|$,  $|\boldsymbol{B}_{DC,min}|$ can be obtained as an average of the crossing points between consecutive $m_F$ levels. The field at which such crossings occur can be obtained analytically by solving equation (\ref{Resonance_condition}), expanding the Zeeman shifted energies (\ref{Zeeman_energy}) up to second order in $\mu_BB/A$. This procedure gives us
\begin{equation}
|\boldsymbol{B}_{DC,min}| = \frac{A}{2F\mu_B} \sum_{m_F=-F}^{m_F=F-1} \frac{g_F - \sqrt{g_F^2 \mp 4(1-2m_F)G_F \frac{\hbar \omega_{RF}}{A}}}{2 (1-2m_F)G_F} 
\label{eq:Br_approx}
\end{equation}
with
\begin{eqnarray}
G_F &=& \mp  \frac{1}{I+1/2} \left( \frac{g_J-g_I}{2(I+1/2)}\right)^2 .
\label{eq:gfs}
\end{eqnarray}
In equations (\ref{eq:Br_approx})-(\ref{eq:gfs}) the upper and lower signs are chosen according to $F=I+J$ and $F=|I-J|$ respectively.

Analytic expressions for the dressed energy can only be obtained in simple cases. For our example of $^{87}$Rb in its ground state, after ignoring couplings between the $F=I+J$ and $F=|I-J|$, the dressed energies of the $F=1$ manifold are: 
\begin{equation}
  E_i(|\boldsymbol{B}_{DC}|) = -\frac{C}{3} + \frac{2\sqrt{C^2 - 3D}}{3} \cos\left(\frac{\theta+\phi_i}{3}\right) + A \left(-\frac{1}{4} + \frac{g_F}{2|g_F|}(I+0.5) \right)
\end{equation}
with $\phi_i = 0, 2\pi, 4\pi$ corresponding to $i=1,2,3$, and
\begin{eqnarray}
  C &=& -(E^{Z,-}_{1}+E^{Z,-}_0+E^{Z,-}_{1}) \nonumber \\
  D &=& E^{Z,-}_{1}E^{Z,-}_{-1} + E^{Z,-}_{1}E^{Z,-}_{0} + E^{Z,-}_{-1}E^{Z,-}_{0} - 2d^2 + \hbar \omega_{RF}(E^{Z,-}_1- E^{Z,-}_{-1} - \hbar \omega_{RF}) \nonumber \\
  E &=& d^2(E^{Z,-}_{1} + E^{Z,-}_{-1}) - E^{Z,-}_{1}E^{Z,-}_{0}E^{Z,-}_{-1} - \hbar \omega_{RF} E^{Z,-}_0(E^{Z,-}_1- E^{Z,-}_{-1} - \hbar \omega_{RF})\nonumber \\
  R &=&  (9CD-27E-2C^3)/54 \nonumber \\
  Q &=& (3D-C^2)/9 \nonumber \\   
  \theta &=& \arccos (R/\sqrt{-Q^3}) ,
\end{eqnarray}
where  $d =  \frac{\mu_BB_{RF}g_F}{2} \langle m_F\pm1|J_x + I_x|m_F\rangle $ and the Zeeman shifted levels are given by equation (\ref{Zeeman_energy}).

\section*{References}
\bibliography{NonLinearZeeman_RFdressing}

\end{document}